\documentclass[preprint,psfig,aps,showpacs]{revtex4}

\usepackage{graphicx}

\begin{document}

\title{Spin structure factors and valence-bond-solid states of the
trimerized Heisenberg chains in a magnetic field}
\author{Shou-Shu Gong, Bo Gu and Gang Su$^{{\ast }}$}
\affiliation{College of Physical Sciences, Graduate University of Chinese Academy of
Sciences, P. O. Box 4588, Beijing 100049, China}

\begin{abstract}
By means of the density matrix renormalization group (DMRG) method, the
static spin structure factors and the magnetization plateaus of the
trimerized Heisenberg ferromagnet-ferromagnet-antiferromagnet and
antiferromagnet-antiferromagnet-ferromagnet spin chains in the presence of a
magnetic field are elaborately studied. It is found that in the plateau
states, the static structure factor with three peaks does not vary with the
external magnetic field as well as the exchange couplings; the spin
correlation function behaves as a perfect sequence and has a simple relation
with the magnetization per site. An approximate wave function for the
plateau states is proposed, and a picture based on the valence-bond-solid
states is presented in order to understand the origin and the total number
of the magnetization plateaus, which are shown to be in agreement with the
DMRG results.
\end{abstract}

\maketitle

In recent years, owing to their unusual magnetic properties, the
low-dimensional quantum spin systems have been extensively investigated both
theoretically and experimentally (see e.g. Ref. \cite{Introduction}). Among
others, the phenomenon of topological quantization of magnetization has
received particular attention (e.g. \cite%
{jump,Hida,Okamoto,OYA,Kitazawa,Honecker,Chen,F-F-AF-AF,Diamond,Ajiro,Ishii,AF-AF-F,Gu1,Gu2,Gu3,Gu4}%
). Recently, the magnetic and thermodynamic properties of the
ferromagnet-ferromagnet-antiferromagnet (F-F-AF) and
antiferromagnet-antiferromagnet-ferromagnet (AF-AF-F) trimerized Heisenberg
spin chains have been systematically explored\cite{Gu1,Gu2}. For both
systems, it have been observed that the topological quantization condition $%
n(S-m)=integer$, with $n$ the period of the ground state determined by the
explicit spatial structure of Hamiltonian, $S$ the spin, and $m$ the
magnetization per site, is only a necessary, but not a sufficient, condition
for the appearance of magnetization plateaus, and the total number of the
plateaus is $2S+1$. The width of plateaus is found to decrease with
increasing the ferromagnetic interaction, and the plateau-non-plateau
transition is of Kosterlitz-Thouless type, while the position of the
plateaus is independent of the magnitude of the exchange couplings\cite{Gu1}%
. The magnetization plateau is smeared out when temperature is increased,
and it appears the systems fall into different thermodynamic states under
different magnetic fields\cite{Gu2}. In order for understanding the
spin-spin correlations in such intriguing systems, and demonstrating why the
total number of the plateaus is $2S+1$, we shall, in this paper, investigate
the static spin structure factor that would be directly measured
experimentally for F-F-AF and AF-AF-F trimerized Heisenberg chains with $%
S=1/2$ and $1$, and offer a picture based on the valence-bond-solid (VBS)
states for further understanding the origin and the total number of the
plateaus.

The Hamiltonian of the trimerized Heisenberg spin chain in a magnetic field
is given by 
\begin{eqnarray}
H &=&\sum\limits_{j}(J\mathbf{S}_{3j-2}\cdot \mathbf{S}_{3j-1}+J\mathbf{S}%
_{3j-1}\cdot \mathbf{S}_{3j}  \nonumber \\
&+&J^{\prime }\mathbf{S}_{3j}\cdot \mathbf{S}_{3j+1})-h\sum%
\limits_{j}S_{j}^{z},  \label{model Hamiltonian}
\end{eqnarray}%
where $J$, $J^{\prime }>0$ denote the AF coupling and $<0$ the F coupling,
and $h$ is the external magnetic field. We take $g\mu _{B}=1$ for
simplicity. Two cases with $S=1/2$ and $1$ will be considered: (a) $%
J=-J_{F}<0$, $J^{\prime }=J_{AF}>0$, the system (F-F-AF) is an
antiferromagnet; (b) $J=J_{AF}>0$, $J^{\prime }=-J_{F}<0$, the system
(AF-AF-F) is a ferrimagnet. The static structure factor $S(q)$ is the
Fourier transform of the spin correlation function $\langle
S_{n}^{z}S_{n^{\prime }}^{z}\rangle $, defined by \cite{note-1} 
\begin{equation}
S(q)=\frac{1}{N}\sum_{n,n^{\prime }}\ e^{iq(n-n^{\prime })}\langle
S_{n}^{z}S_{n^{\prime }}^{z}\rangle .  \label{s(q)}
\end{equation}%
The density matrix renormalization group (DMRG) method \cite{DMRG1,DMRG2}
will be invoked. In the following calculations, the length of the chain is
taken as $N=60$, and the number of the optimal states is kept as $M=60$.
Open boundary conditions are adopted. The truncation error is less than $%
10^{-3}$ in all calculations.

\begin{figure}[tbp]
\includegraphics[width = 1.0\linewidth,clip]{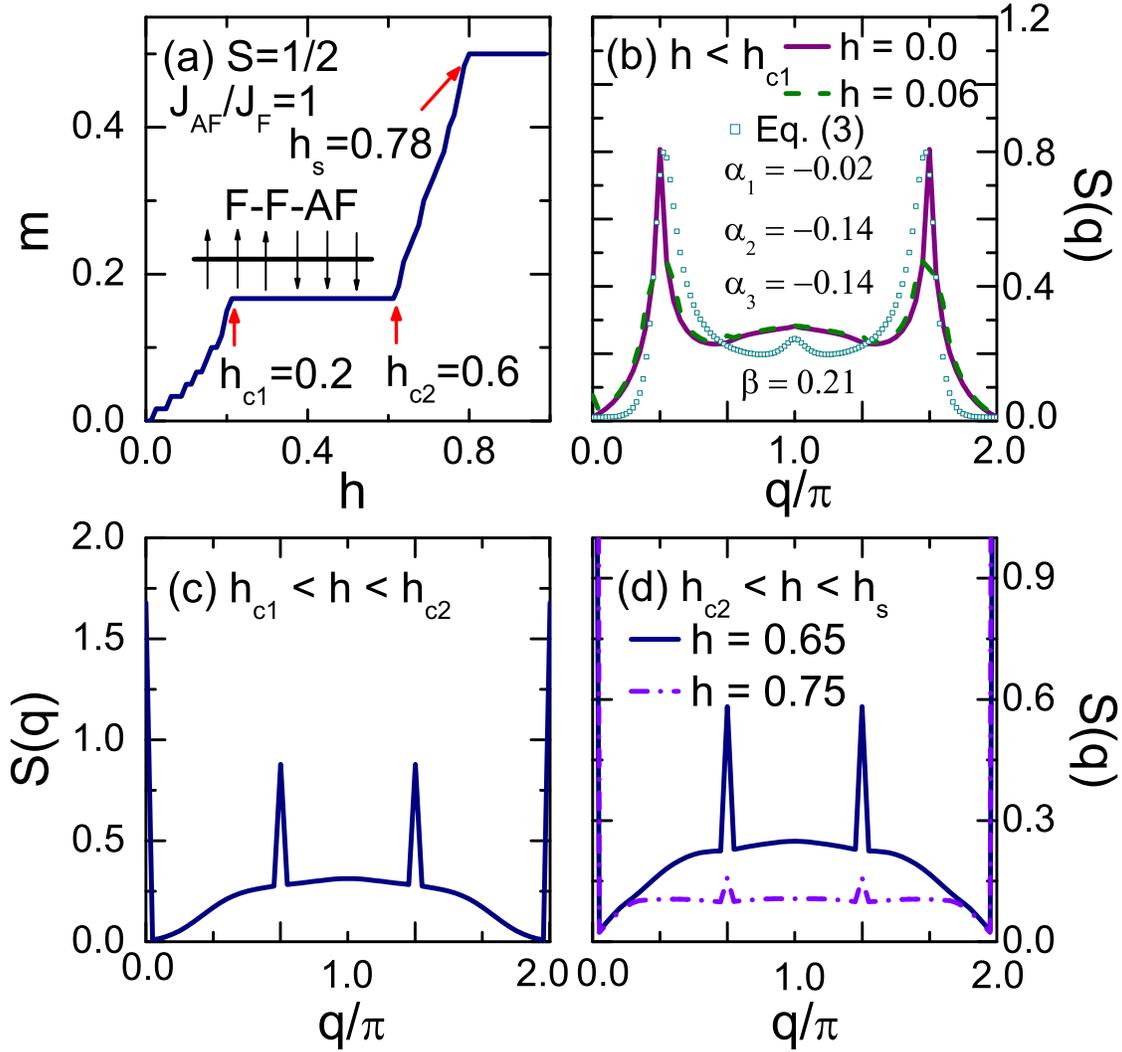}
\caption{(Color online) For a spin-$1/2$ F-F-AF trimerized Heisenberg chain
with couplings $J_{AF}/J_{F}=1$, (a) the magnetization per site $m$ as a
function of the external magnetic field $h$ in the ground states; and the
static structure factor $S(q)$ in the ground states under the external field
(b) $h<h_{c1}$, (c) $h_{c1}<h<h_{c2}$, and (d) $h_{c2}<h<h_{s}$.}
\label{ffahalf}
\end{figure}

For the spin-$1/2$ F-F-AF chain with couplings $J_{AF}/J_{F}=1$, one
magnetization plateau with $m=1/6$ is observed, being consistent with the
quantization condition\cite{OYA}. Let $h_{c1}$ and $h_{c2}$ be the lower and
upper critical fields at which the magnetization plateau appears and
disappears, respectively, and $h_{s}$ be the saturation field, as indicated
in Fig. \ref{ffahalf}(a). When $h<h_{c1}$, as demonstrated in Fig. \ref%
{ffahalf}(b), the static spin structure factor $S(q)$ shows three peaks at $%
q=\pi /3$, $5\pi /3$, and $\pi $. With increasing the magnetic field, the
peak at $q=\pi $ disappears while the peak at $q=\pi /3$ ($5\pi /3$) splits
into two small peaks that continuously shift towards $q=0$ and $2\pi /3$ ($%
q=0$ and $4\pi /3$), respectively. Meanwhile, new peaks start to appear in $%
q=0$, $2\pi /3$ and $4\pi /3$. When $h_{c1}<h<h_{c2}$, where the ground
states are in the plateau state with $m=1/6$, as shown in Fig. \ref{ffahalf}%
(c), $S(q)$ with three peaks at $q=0$, $2\pi /3$ and $4\pi /3$ does not vary
with the magnetic field, which corresponds to the long-range order and a
perfect sequence for the spin correlation function $\langle
S_{0}^{z}S_{j}^{z}\rangle $, which we will discuss later. When $%
h_{c2}<h<h_{s}$, the peaks at $q=2\pi /3$ and $4\pi /3$ are suppressed with
the increasing field and disappear at the saturation field, while the peak
at $q=0$ increases with the field, which show the ferromagnetic
characteristic, as demonstrated in Fig. \ref{ffahalf}(d).

In the absence of the external field, we find that the static structure
factor $S(q)$ could be fitted by 
\begin{eqnarray}
S(q) &=&\sum_{l=1}^{3}\{\alpha _{l}e^{-l\beta }[\frac{2\sin 3q(\cosh 3\beta
-\cos 3q)\sin lq}{\cosh (6\beta )-\cos (6q)}  \nonumber \\
&+&\frac{(e^{6\beta }-\cos 6q-2\cos 3q\sinh 3\beta )\cos lq}{\cosh (6\beta
)-\cos (6q)}]\}  \nonumber \\
&+&\frac{S(S+1)}{3},  \label{threemodel}
\end{eqnarray}%
where $\alpha _{l}$ and $\beta $ are parameters. As presented in Fig. \ref%
{ffahalf}(b), the characteristic peaks can be well fitted by Eq. (\ref%
{threemodel}), with only a slightly quantitative derivation, showing that
the main features of $S(q)$ of this model can be captured by a linear
superposition of three modes, which is closely related to the low-lying
excitations of these trimerized spin systems\cite{Gu1,Gu4}.

In the plateau states, $S(q)$ has three peaks at $q=0$, $2\pi /3$ and $4\pi
/3$, and $\langle S_{0}^{z}S_{j}^{z}\rangle $ shows a perfect sequence with
a period of $3$. It is the period of $\langle S_{0}^{z}S_{j}^{z}\rangle $
that determines the peak positions of $S(q)$. Define $\xi
(m)=(1/N)\sum_{n}\langle S_{n}^{z}S_{n-m}^{z}\rangle $, which also shows a
perfect sequence with a period of $3$, say, $\{\cdots ,(\xi (3j),\xi
(3j+1),\xi (3j+2)),\cdots \}$. Thus, the structure factor $S(q)$ in the
plateau states can be written as 
\begin{eqnarray}
S(q) &=&\frac{1}{3}S(S+1)+2\xi (1)\cos (q)+2\xi (2)\cos (2q)  \nonumber \\
&+&2\sum_{j=1}^{\infty }\xi (3j)\cos (3jq)+2\sum_{j=1}^{\infty }\xi
(3j+1)\cos (3j+1)q  \nonumber \\
&+&2\sum_{j=1}^{\infty }\xi (3j+2)\cos (3j+2)q.  \label{Sq}
\end{eqnarray}%
Because the series $\sum_{j=1}^{\infty }\cos (3jq)$ only diverges at $q=0$, $%
2\pi /3$ and $4\pi /3$, and oscillates at other $q$, the peak positions of $%
S(q)$ appear only at $q=0$, $2\pi /3$ and $4\pi /3$, consistent with our
numerical calculations. This argumentation may be generalized to other spin
chains, say, if $\langle S_{0}^{z}S_{j}^{z}\rangle $ behaves as a perfect
sequence with a period of $n$, $S(q)$ would only have peaks at $q=0$, $2\pi
/n$, $4\pi /n$, $\cdots $.

Besides the sequence structure of the spin correlation function, it is found
that, from our DMRG results, in the plateau states the spin correlation
functions satisfy the following relation: 
\begin{equation}
\frac{1}{3}[\xi (3j)+\xi (3j+1)+\xi (3j+2)]=m^{2},  \label{candm}
\end{equation}
which does not alter with the couplings $J_{AF}/J_{F}$ as well as the
magnitude of the spin $S$. As the susceptibility $\chi (T)=\frac{1}{k_{B}T}%
\frac{1}{N}\sum_{i,j}[\langle S_{i}^{z}S_{j}^{z}\rangle -\langle
S_{i}^{z}\rangle \langle S_{j}^{z}\rangle ]$ bears a finite value at $%
T\rightarrow 0$ in the plateau states because of $m(T)$ being a constant, it
leads to 
\begin{equation}
\frac{1}{N}\sum_{i,j}[\langle S_{i}^{z}S_{j}^{z}\rangle -\langle
S_{i}^{z}\rangle \langle S_{j}^{z}\rangle ]=0  \label{generalresult}
\end{equation}%
at $T\rightarrow 0$. Since $\langle S_{0}^{z}S_{j}^{z}\rangle $ shows a
perfect sequence in the plateau states, we can get Eq. (\ref{candm})
directly from Eq. (\ref{generalresult}).

The spin-$1$ F-F-AF Heisenberg chain with the couplings $J_{AF}/J_{F}=1$ is
also studied. The two plateaus at $m=1/3$ and $2/3$ are observed. The static
structure factor and spin correlation function show the same qualitative
characteristics as those of the spin-$1/2$ chain. The zero-field static
structure factor $S(q)$ can also be well fitted by Eq. (\ref{threemodel}),
and in both plateaus, Eq. (\ref{candm}) is still satisfied.

\begin{figure}[tbp]
\includegraphics[width = 1.0\linewidth,clip]{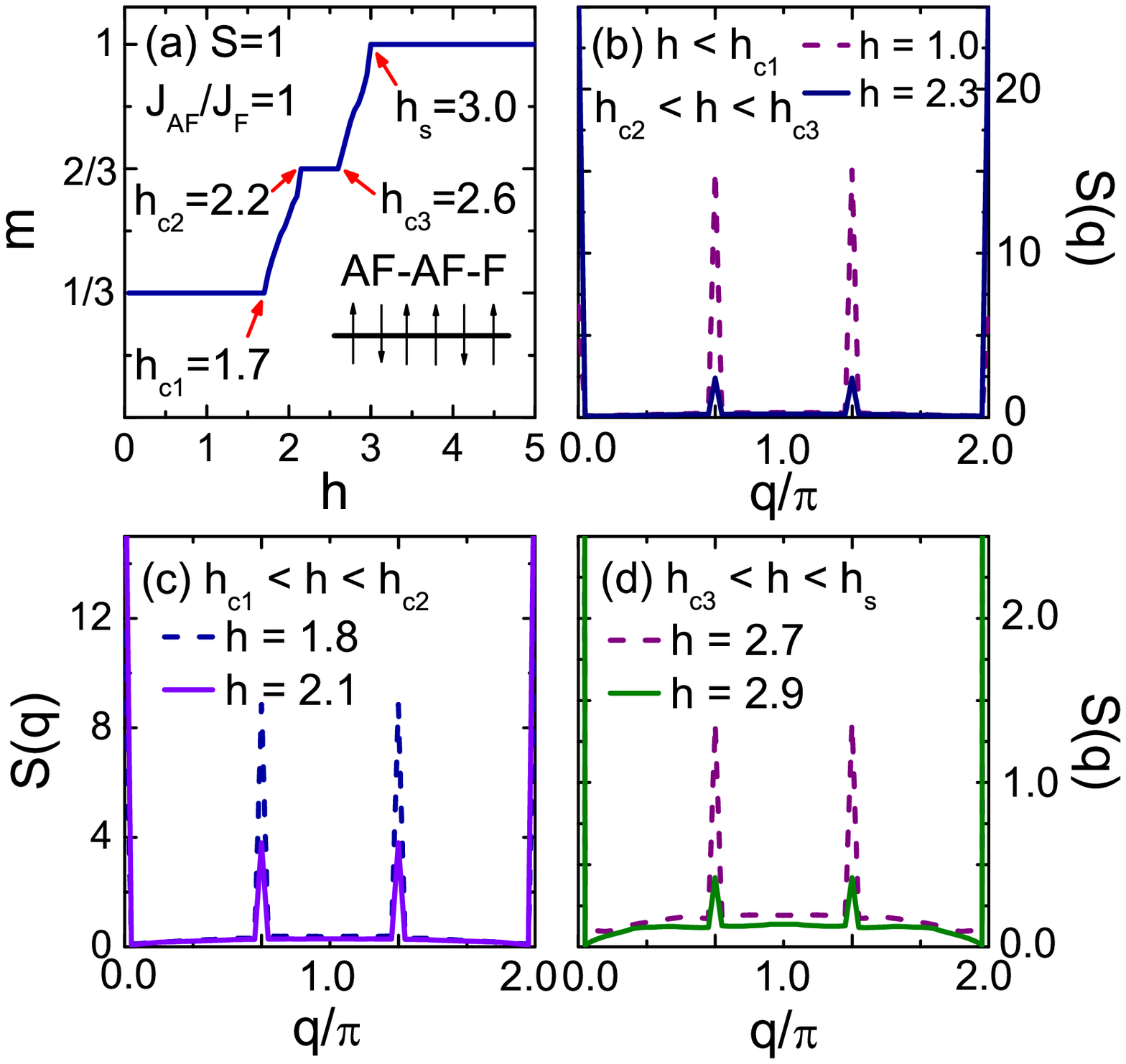}
\caption{(Color online) For a spin-$1$ AF-AF-F trimerized Heisenberg chain
with fixed couplings $J_{AF}/J_{F}=1$, (a) the magnetization per site $m$ as
a function of the external magnetic field $h$ in the ground states; and the
static structure factor $S(q)$ in the ground states with the external field
(b) $h<h_{c1}$ and $h_{c2}<h<h_{c3}$, (c) $h_{c1}<h<h_{c2}$, (d) $%
h_{c3}<h<h_{s}$. }
\label{aafone}
\end{figure}

The AF-AF-F trimerized Heisenberg chains with spin $1/2$ and $1$ are
investigated as well. For the spin-$1$ AF-AF-F Heisenberg chain with
couplings $J_{AF}/J_{F}=1$, two plateaus at $m=1/3$ and $m=2/3$ are
observed. Let $h_{c1}$ denote the critical field of disappearance of $m=1/3$
plateau, $h_{c2}$ and $h_{c3}$ be the critical fields of appearance and
disappearance of $m=2/3$ plateau, respectively, and $h_{s}$ be the
saturation field, as marked in Fig. \ref{aafone}(a). When no magnetic field
is applied, $S(q)$ shows three peaks at $q=0$, $2\pi /3$ and $4\pi /3$,
which differs from the case of the F-F-AF chain because the latter is an
antiferromagnet. When $h<h_{c1}$ and $h_{c2}<h<h_{c3}$, where the ground
states are in $m=1/3$ and $2/3$ plateau states, $S(q)$ with three peaks at $%
q=0$, $2\pi /3$ and $4\pi /3$ does not vary with the external field, both of
which show a perfect sequence with a period of $3$ for $\langle
S_{0}^{z}S_{j}^{z}\rangle $, as shown in Fig. \ref{aafone}(b). When $%
h_{c3}<h<h_{s}$, like the case of the F-F-AF chain, the peaks at $q=2\pi /3$
and $4\pi /3$ are depressed with increasing the field, and disappear at the
saturation field, while the peak at $q=0$ increases with the field, as shown
in Fig. \ref{aafone}(d).

The spin-$1/2$ AF-AF-F Heisenberg chain with couplings $J_{AF}/J_{F}=1$ is
also explored. The features similar to those of $S=1$ are seen, and the spin
correlation functions also satisfy Eq. (\ref{candm}) in the plateau states.

The plateau states for the F-F-AF and AF-AF-F trimerized Heisenberg chains
with spin $S$ can be described by an approximate wave function defined by 
\begin{eqnarray}
\psi _{i} &=&a|\uparrow _{3i-2}\uparrow _{3i-1}\downarrow _{3i}\rangle
+b|\uparrow _{3i-2}\downarrow _{3i-1}\uparrow _{3i}\rangle  \nonumber \\
&+&c|\downarrow _{3i-2}\uparrow _{3i-1}\uparrow _{3i}\rangle ,\quad
(i=1,\cdots ,N/3)  \label{groundstate}
\end{eqnarray}%
where $\uparrow _{j}$ $(\downarrow _{j})$ denotes spin up (down) on site $j$%
. It is found that, if the coefficients $a$, $b$ and $c$ are properly
chosen, one could find the so-calculated local magnetic moment $\langle
S_{j}^{z}\rangle $ as well as the spin correlation functions to be in
agreement with the DMRG results. Take the spin-$1/2$ F-F-AF trimerized
Heisenberg chain for example. In the plateau state with $m=1/6$, the DMRG
results show that $\langle S_{j}^{z}\rangle $ behaves as $\{\cdots
,(S_{1},S_{2},S_{2}),\cdots \}$ with $S_{1}=0.367$ and $S_{2}=0.067$, giving
rise to the magnetization per site $m=(S_{1}+2S_{2})/3=1/6$, and the spin
correlation function $\xi (j) $ behaves such that $\{\cdots ,(\xi (3j),\xi
(3j+1),\xi (3j+2)),\cdots \}$ with $\xi (3j)=0.048$ and $\xi (3j+1)=\xi
(3j+2)=0.018$, satisfying $\frac{1}{3}[\xi (3j)+\xi (3j+1)+\xi (3j+2)]=(%
\frac{1}{6})^{2}$. If we choose the coefficients $a=\pm \sqrt{13/30}$, $%
b=\pm \sqrt{13/30}$ and $c=\pm \sqrt{2/15}$, from the approximate wave
function we can obtain $\langle \psi _{i}|S_{3i-2}^{z}|\psi _{i}\rangle
=11/30$, $\langle \psi _{i}|S_{3i-1}^{z}|\psi _{i}\rangle =\langle \psi
_{i}|S_{3i}^{z}|\psi _{i}\rangle =1/15$, and $\xi (3j)=43/900$, $\xi
(3j+1)=\xi (3j+2)=4/225$, both of which are quite in agreement with our DMRG
results.

\begin{figure}[tbp]
\includegraphics[width = 0.75\linewidth,clip]{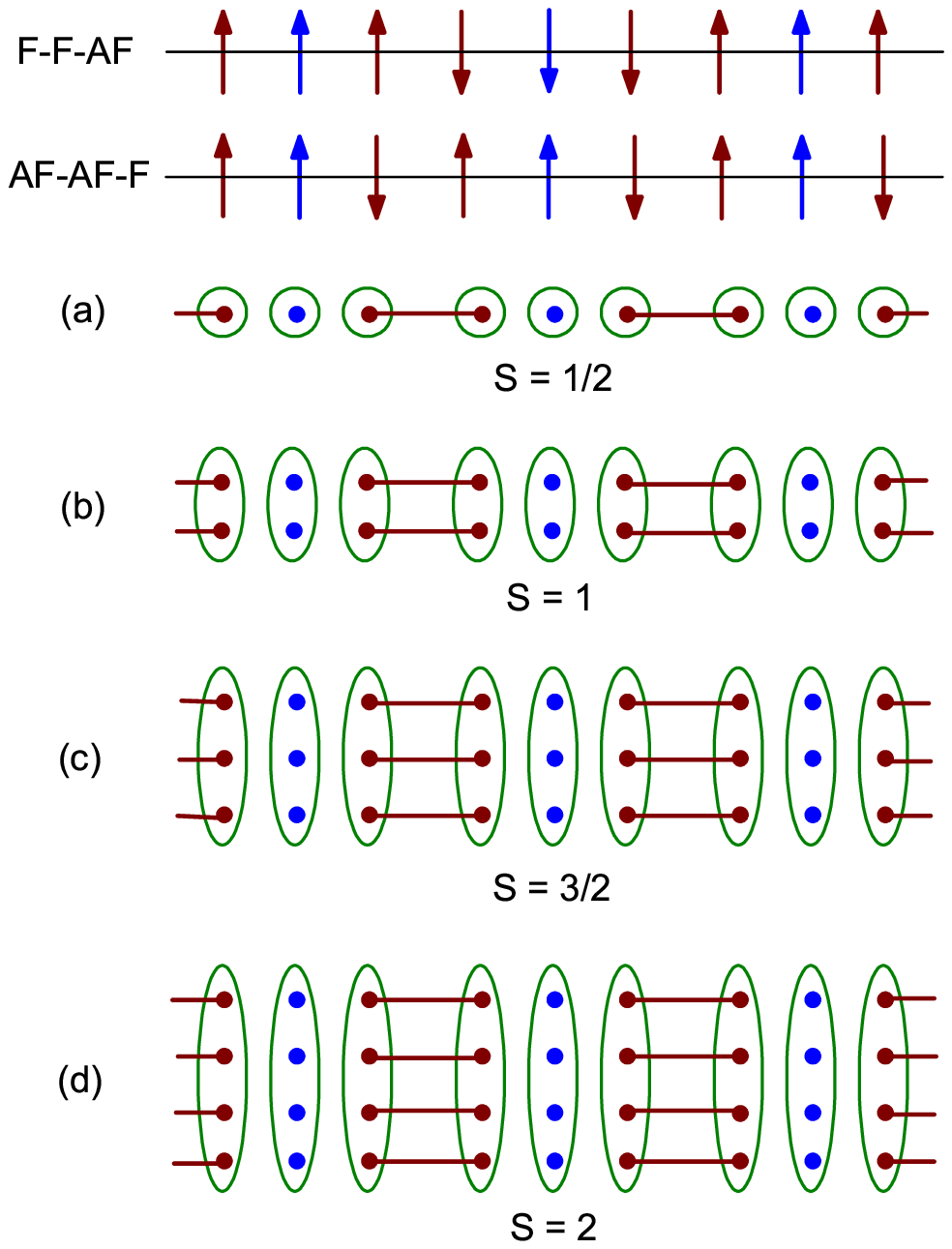}
\caption{(Color online) Sketch of the VBS picture of the F-F-AF and AF-AF-F
trimerized Heisenberg chains with spin (a) $S=1/2$, (b) $1$, (c) $3/2$, and
(d) $2$, respectively. $\uparrow $ denotes spin up, and $\downarrow $ spin
down. The bullet $\bullet $ denotes a spin $1/2$, and the short line between
bullets denotes the spin singlet state. The big open elipse presents spin $S$
per site which could be viewed as a combination of $2S$ spin $1/2$.}
\label{vbs}
\end{figure}

Let us now invoke the VBS picture to show why the total number of the
magnetization plateaus in these trimerized systems is $2S+1$. As shown in
Fig. \ref{vbs}, we denote a spin $1/2$ by a bullet, a spin singlet of two
spin $1/2$ by a short line between two bullets, and the spin $S$ per site by
a large open ellipse that could be viewed as a combination of $2S$ spin $1/2$%
. For the F-F-AF chains with spin $S=1/2$, $1$, $3/2$ and $2$, the isolated
spins $S$ that are denoted by large open ellipses, are separated by spin
singlet states. Thus, these isolated spins could be regarded as a classic
spin chain with antiferromagnetic couplings, whose magnetization curves
should have no plateau except for the saturation plateau of $m=S/3$. This
picture explains why the magnetization plateau of $m<S/3$, which is
permitted by the quantization condition\cite{OYA}, does not appear. For the
ferrimagnetic AF-AF-F chains with $S=1/2$, $1$, $3/2$ and $2$, the isolated
spins are aligned parallel, and the magnetization per site in the absence of
the magnetic field is $S/3$, leading to the plateaus of $m<S/3$ never exist.
With increasing the magnetic field, the spin singlet bonds are gradually
broken. When a spin singlet bond per three sites is simultaneously broken, a
new magnetization plateau appears, and so on. When $2S$ spin singlet bonds
per three sites are simultaneously broken, there are $2S$ plateaus
occurring. Considering that there is also a saturation plateau, the
magnetization per site obeys $m=(S+x)/3$ $(x=0, 1, \cdots, 2S)$. Therefore,
the total number of magnetization plateaus is $2S+1$, which is well in
agreement with the DMRG results\cite{Gu1}. For instance, the trimerized
Heisenberg chain with $S=2$, according to our VBS picture, exhibits
magnetization plateaus at $m=2/3$, $1$, $4/3$, $5/3$ and $2$, which have
been confirmed by the DMRG results\cite{Gu1}.

In summary, we have studied the static structure factor $S(q)$ and the
plateau states of the F-F-AF and AF-AF-F trimerized Heisenberg chains. It
has been found that in plateau states, $S(q)$ with three peaks at $q=0$, $%
2\pi /3$ and $4\pi /3$ does not vary with the magnetic field as well as the
exchange couplings $J_{AF}/J_{F}$, and the peak positions are determined by
the period of the spin correlation function. From the DMRG data, it is
observed that the spin correlation function satisfies a simple relation with
the magnetization per site [Eq.(\ref{candm})]. An approximate trimerized
state is presented to describe the physical properties in the plateau
states, and the results obtained by using this trial wave function are quite
in agreement with the DMRG data. To explain the origin and the total number
of the plateaus, a picture in terms of the VBS states is offered. From this
picture, the appearance of the magnetization plateaus can be attributed to
the breaking of the spin singlet states and the isolated spins. The total
number and the positions of the plateaus deduced from our VBS picture are
exactly the same with the DMRG results. It is expected that the properties
of the static structure factors reported in this paper could be useful for
future experimental studies on the trimerized Heisenberg spin chain
compounds.

\acknowledgments

We are grateful to X. Chen, Z. C. Wang, Q. B. Yan and Q. R. Zheng for useful
discussions. This work is supported in part by the National Science Fund for
Distinguished Young Scholars of China (Grant No. 10625419), the National
Science Foundation of China (Grant Nos. 90403036 and 20490210), and by the
MOST of China (Grant No. 2006CB601102).


\begin{thebibliography}{99}
\bibitem[]{Permanent address.} $^{\ast }$Corresponding author. E-mail:
gsu@gucas.ac.cn

\bibitem{Introduction} H. -J. Mikeska and A. K. Kolezhuk, \textit{%
One-Dimensional Magnetism}, Lecture Notes in Physics Vol. 645, edited by U.
Schollwock, J. Richter, D. J. J. Farnell and R.F. Bishop (Springer-Verlag,
Berlin, Heidelberg, 2004).

\bibitem{jump} J. Schulenburg, A. Honecker, J. Schnack, J. Richter, and H.
J. Schmidt, \textit{Phys. Rev. Lett.} \textbf{88}, 167207 (2002); J.
Schnack, H. J. Schmidt, A. Honecker, J. Schulenberg and J. Richter, \textit{%
J. Phys. Condens. Matter} \textbf{51}, 43 (2006).

\bibitem{Hida} K. Hida, \textit{J. Phys. Soc. Jpn.} \textbf{63}, 2359 (1994).

\bibitem{Okamoto} K. Okamoto, \textit{Solid State Commun.} \textbf{98}, 245
(1996).

\bibitem{OYA} M. Oshikawa, M. Yamanaka and I. Affleck, \textit{Phys. Rev.
Lett.} \textbf{78}, 1984 (1997).

\bibitem{Kitazawa} A. Kitazawa and K. Okamoto, \textit{J. Phys. Condens.
Matter} \textbf{11}, 9765 (1999).

\bibitem{Honecker} A. Honecker, \textit{Phys. Rev. B} \textbf{59}, 6790
(1999).

\bibitem{Chen} W. Chen, K. Hida and B. C. Sanctuary, \textit{J. Phys. Soc.
Jpn.} \textbf{69}, 3414 (2000); W. Chen, K. Hida and B. C. Sanctuary, 
\textit{Phys. Rev. B} \textbf{63}, 134427 (2001).

\bibitem{F-F-AF-AF} J. Stre\v{c}ka, M. Ja\v{s}\v{c}ur, M. Hagiwara, Y.
Narumi, K. Kindo, and K. Minami, \textit{Phys. Rev. B} \textbf{72}, 024459
(2005).

\bibitem{Diamond} H. Kikuchi, Y. Fujii, M. Chiba, S. Mitsudo, T. Idehara, T.
Tonegawa, K. Okamoto, T. Sakai, T. Kuwai and H. Ohta, \textit{Phys. Rev.
Lett.} \textbf{94}, 227201 (2005).

\bibitem{Ajiro} Y. Ajiro, T. Asano, T. Inami, H. Aruga-Katori and T. Goto, 
\textit{J. Phys. Soc. Jpn.} \textbf{63}, 859 (1994).

\bibitem{Ishii} M. Ishii, H. Tanaka, M. Hori, H. Uekusa, Y. Ohashi, K.
Tatani, Y. Narumi and K. Kindo, \textit{J. Phys. Soc. Jpn.} \textbf{69}, 340
(2000).

\bibitem{AF-AF-F} M. A. M. Abu-Youssef, M. Drillon, A. Escure, M. A. S.
Goher, F. A. Mautner and R. Vicente, \textit{Inog. Chem.} \textbf{39}, 5022
(2000).

\bibitem{Gu1} B. Gu, G. Su and S. Gao, \textit{J. Phys. Condens. Matter} 
\textbf{17}, 6081 (2005).

\bibitem{Gu2} B. Gu, G. Su and S. Gao, \textit{Phys. Rev. B} \textbf{73},
134427 (2006).

\bibitem{Gu3} B. Gu and G. Su, \textit{Phys. Rev. Lett.} \textbf{97}, 089701
(2006).

\bibitem{Gu4} B. Gu and G. Su, \textit{Phys. Rev. B} \textbf{75}, 174437
(2007).

\bibitem{note-1} As the lattice of the system can be devided into three
sublattices, the static structure factor $S(q)$ could be obtained in terms of
a $3\times 3$ matrix. Our calculations show that, although the sublattice
structure factor exhibits different behaviors, the total structure factor $S(q)$ 
obtained by a $3\times 3$ matrix gives exactly the same result 
as that calculated directly by Eq. (\ref {s(q)}).

\bibitem{DMRG1} S. R. White, \textit{Phys. Rev. Lett.} \textbf{69}, 2863
(1992); S. R. White, \textit{Phys. Rev. B} \textbf{48}, 10345 (1993).

\bibitem{DMRG2} U. Schollw\"{o}ck, \textit{Rev. Mod. Phys.} \textbf{77}, 259
(2005).
\end{thebibliography}
\end{document}